\magnification=\magstep1
\def\giorno{27/1/97}

\def\G{{\cal G}}
\def\A{{\cal A}}
\def\B{{\cal B}}
\def\C{{\cal C}}
\def\F{{\cal F}}
\def\L{{\cal L}}
\def\M{{\cal M}}
\def\R{{\cal R}}
\def\U{{\cal U}}
\def\V{{\cal V}}

\def\S{\Sigma}
\def\Ga{\Gamma}
\def\ga{\gamma}
\def\d{{\rm d}}
\def\de{\delta}
\def\eps{\varepsilon}
\def\om{\omega}

\def\la{\lambda}
\def\a{\alpha}
\def\b{\beta}
\def\pa{\partial}
\def\T{{\rm T}}
\def\sse{\subseteq}
\def\ss{\subset}
\def\<{\langle}
\def\>{\rangle}

\def\D{\nabla}
\def\grad{\nabla}

\def\({\left(}
\def\){\right)}
\def\[{\left[}
\def\]{\right]}

\def\ref#1{[#1]}
\def\~#1{{\widetilde #1}}
\def\=#1{{\bar #1}}

\font\petitrm =  cmr9 
\font\petitit = cmsl9
\font\petitbf = cmbx9

\def\section*#1{\bigskip \bigskip {\bf #1} \bigskip}

{\nopagenumbers \parindent=0pt
~\vskip 2 truecm
\centerline{\bf Michel theory of symmetry breaking}
\medskip
\centerline{\bf and gauge theories}
\footnote{}{{\tt  \giorno }}

\footnote{}{{\petitit Research supported by the Volkswagen
Stiftung under the RiP program at Oberwolfach.}}
\bigskip\bigskip\bigskip

\vfill
\centerline{Giuseppe Gaeta$^*$} 
\medskip
\centerline{\it I.H.E.S., 35 Route de Chartres}
\centerline{\it 91440 Bures sur Yvette (France)}
\centerline{\tt gaeta@ihes.fr}
\bigskip
\bigskip
\bigskip
\centerline{Paola Morando}
\medskip 
\centerline{\it Dipartimento di Matematica, Politecnico di Torino,}
\centerline{\it Corso Duca degli Abruzzi 24, 10129 Torino (Italy)}
\centerline{\tt morando@polito.it}

\vfill

{\bf Summary.} {We extend Michel's theorem on the geometry of symmetry
breaking \ref{1} to the case of pure gauge theories, i.e. of
gauge-invariant functionals defined on the space $\C$ of connections of a
principal fiber bundle. Our proof follows closely the original one by
Michel, using several known results on the geometry of $\C$. The result
(and proof) is also extended to the case of gauge theories with matter
fields.}
\bigskip\bigskip

\vfill 
{\petitrm $^*$ On leave from Dept. of Mathematics, Loughborough University,
Loughborough LE11 3TU (GB)}
\eject}
\pageno=1

\parskip=10pt
\parindent=0pt

\section*{Introduction}

In 1971, Louis Michel -- motivated by the $SU(3)$ theory of hadronic
interactions -- proved a remarkable theorem on symmetry breaking in theories
described by a $G$-invariant potential ($G$ a compact semisimple Lie group)
over a finite dimensional smooth $G$-manifold $M$ \ref{1}; this result was
a direct generalization of the theory he and Radicati had developed to study
the geometry of the $SU(3)$ octet, and the model-independent features of the
$SU(3)$ theory \ref{2-4}.

Essentially, Michel theorem guarantees that, under suitable conditions,
there are points in $M$ which are critical for {\it any} $G$-invariant
potential $V: M \to \R$; moreover, these points are characterized in terms
of a geometric construction which takes into account the symmetry properties
of points -- and subsets -- of $M$ under the $G$ action. Notice that
actually, critical points of $G$-invariant functions come necessarily in
$G$-orbits; thus, whenever the orbit space $\Omega = M/G$ is well defined --
as it is under the conditions mentioned above -- it is convenient to set the
problem directly in $\Omega$ (maybe with an explict use of the basis of
$G$-invariant functions on $M$ as coordinates in $\Omega$), as Michel did.

Thus, in summary, the Michel theorem on critical orbits of $G$-invariant
functions on a $G$-manifold \ref{1} allows to identify $G$-orbits which
are critical for {\it any} $G$-invariant function; thus, it permits to {\it
study spontaneous symmetry breaking in a model-independent way} 
\ref{3-16}; 
this theory has also been extended to study
supersymmetry breaking \ref{12}. Mathematical foundations for the theory are
provided e.g. by \ref{17,18}.

The purpose of the present note is to show that Michel's theory can
be extended to the study of (pure) gauge theories, such as those defined by
a Yang-Mills functional, and more generally by a gauge-invariant functional
defined on a space of connections for a given principal bundle.

It will turn out that once this is obtained, extension to the case of complete
gauge theories, i.e. theories with matter fields, follows rather
easily\footnote{$^{1}$}{In this
respect, it should be mentioned that an extension of Michel theory to gauge
functionals (in that case, dealing only with matter fields) was already considered in
\ref{19-21}, but this suffered from several -- strong -- limitations; these
followed from focusing on the seemingly easier sector of matter fields, and it turns
out that the present approach is not only more natural geometrically, but also more
fruitful.}.

It should be stressed that, for our result to be of physical interest, the
space of connections to be considered (see below) should obey some natural
conditions.  Thus, on the one side we would be not justified in restricting
our attention to the (dense) subspace of irreducible connections, for which
a full geometric description is well known \ref{22,23}, and which is
relatively simple (we recall however that in the $SU(2)$ case
this coincides with the full space of connections \ref{24}). On the other
side, we should {\it not} consider the full space of connections, but only
those connections satisfying a {\it finite energy condition}, or more
precisely the completion of this set in suitable norm; this leads naturally
to consider Sobolev  spaces of connections and Sobolev norms \ref{24,25}.
The same remark would apply when considering sections of a vector bundle
in the context of gauge theories with matter fields, see below.

In the present note, we will give a careful statement and a complete proof
of the extension of Michel's theorem to pure gauge theories; we will
also shortly discuss some -- easy but relevant -- extensions and
generalizations of this (including the one to theories with matter fields), 
whose proof is only sketched or omitted at all as
it would just repeat the one for the pure gauge case, and some of the
possible applications.

It is appropriate to stress here that it turns out -- as it will be clear
in our  discussion -- that one can directly extend the essence of Michel's
construction and  proof, from its original setting -- $M$ a finite dimensional
manifold, $G$ a compact Lie group, in the following referred to as the
``classical case'' -- to the case of gauge theories, which involve infinite dimensional,
non-compact manifold and group (a large part of the results
needed for this are classical ones \ref{22,23,26}); this shows (see also the final
discussion in sect.7) how far-reaching is the Michel theory of symmetry breaking.

{\it Plan of the paper.} Let us shortly describe the pan of the paper. 
We will start by collecting in sect.1 some well known 
geometrical fact
to provide a proper setting -- and fix notation -- for the discussion to follow.
In sect.2 we will define a Sobolev norm on the space of connections and on gauge
sections, and choose some Sobolev class of connections -- and correspondingly of
gauge sections -- in which we work; this is needed in order to deal with Hilbert
spaces. With this, we can use known results concerning existence of slices. The
central result of this note is given in sect.3; it is a direct extension of Michel's
theorem \ref{1} -- and proof -- to the pure gauge case; after this, we can pass to
consider extension and consequences of it. 
Thus, first of all we notice in sect.4 that
once we have the result for the pure gauge case, it can be easily extended to the case
of gauge theories with matter fields. Then, in sect.5 we discuss some other
extensions, dealing with general systems (i.e. not necessarily variational ones) with
a gauge symmetry, and we formulate the gauge equivalent of reduction lemmata which are 
well known in the finite dimensional case; these are applied in sect.6 to discuss
specifically the problem of symmetry breaking in theories with a (continuous)
dependence on a parameter. Finally, in sect.7 we discuss some further extensions, as
well as the strength -- and weakness -- of the approach proposed here. The discussion
of sects.5-7 refers only to the pure gauge case for ease of notation, 
but it will be clear from sect.4 that these would also extend to the case where matter
fields are present.
All our discussion will be conducted using only the geometry of the problem, i.e. with
no reference to the kind of specific functional one encounters in physical examples of
gauge theories. To discuss these is obviously a very relevant matter, so that we do
indeed consider them, and how our main theorem can be of use in determining specific
solutions, in the appendices. Here we restrict specifically to the well known class of
functionals made of the Yang-Mills one $L = |F_{\mu \nu } |^2$ for the pure gauge part, 
and by the standard
Lagrangian $L = [ | \D \phi |^2 - V (\phi) ]$ (see below) for the matter fields (with
$\grad$ the covariant derivative). In appendix A we study of special class of
solutions, in appendix B we further restrict this to the case associated to the null
connection $A^0$ (see below), and finally in appendix C we deal with the case
$G=SU(3)$.

\section*{1. Geometrical setting}

Let $B$ be an $n$-dimensional ($n$ finite) riemannian manifold (e.g. $B=
R^4$ or $S^4$); let $G$ be a compact, connected and semisimple Lie group
(e.g. $G = SU(N)$); let $P$ be
a principal fiber bundle over $B$ with projection $\pi : P \to B$ and fiber
$\pi^{-1}(x) \approx G$. We denote by $\G$ the Lie algebra of $G$.

Let $\C$ be the set of connections on $P$. It is well known that $\C$ is an
affine space modelled on the vector space $\A := \Lambda^1 \( \T^*B,\G \)$
of the one-forms on $B$ taking value in $\G$. With any connection $A \in \C$
we can associate a covariant derivative $\D^\a$ and a one-form $\a \in \A$
that are given, in local coordinates $(x^1 , ... , x^n )$ on $B$ and with
$A_\mu : B \to \G$, by $$ \a \ = \ A_\mu (x) \ \d x^\mu \ ; \eqno(1) $$
$$ \D^\a_\mu \ = \ \pa_\mu \ + \ A_\mu \ . \eqno(2) $$

Choosing as reference point in $\C$ the null connection $A^0$ such that
$A^0_\mu = 0$ ($\mu = 1,..., n$), from now on we will say ``the connection
$\a$'' to mean ``the connection $A=A^0+\a$ whose associated connection form
is $\a$''.

Let $\Ga$ denote the space of differentiable sections of $P$; $\Ga$ has the
structure of a Hilbert-Lie group, and a section $\ga \in \Ga$ will be
written in local coordinates as $$ \ga \ = \ g(x) \ , \eqno(3) $$  with $g :
B \to G$. We say that $\Ga$ is a gauge group modelled on $G$.

The gauge group $\Ga$ acts naturally on the operator of covariant derivative
associated to a connection $A \in \C$ by conjugation, i.e. $\ga : \D^\a \to
\ga \D^\a \ga^{-1}$; the action of $\Ga$ on $\A$ is given in local
coordinates by $$ \ga \ : \ A_\mu \ \Longrightarrow \ g(x) \cdot A_\mu (x)
\cdot g^{-1} (x) \ - \ \( \pa_\mu g \) (x) \cdot g^{-1} (x) \ \ . \eqno(4) $$
The action of $\ga \in \Ga$ on the connection form $\a \in \A$ will be
denoted by $\ga (\a )$.

With any connection form $\a$  (i.e. with any connection $A$) we can
associate a {\it gauge isotropy subgroup} $\Ga_\a$, $$ \Ga_\a \ = \ \{ \ga
\in \Ga \ : \ \ga (\a ) = \a \} \ \ . \eqno(5) $$  It is well known
\ref{22} that, with $p_0 \in P$ any reference point in $P$ and $x_0 = \pi
(p_0 )$, $$ \Ga_\a \ = \ \{ \ga \in \Ga \ : \ \D^\a (\ga ) = 0 \ , \ g (x_0
) \in C_G \[ H_\a (p_0 ) \] \ \} \ \ , \eqno(6) $$  where $H_\a (p_0 )$ is
the holonomy group of the connection $\D^\a$ at $p_0$, and $C_G [H]$ the
centralizer of $H$ in $G$, i.e. $$ C_G [H] \ = \ \{ g \in G \ : \ [g,h] = 0
\ \ \forall h \in H \} \ \ . \eqno(7) $$ Thus, $\Ga_\a$ is isomorphic to a
subgroup of the compact group $G$; notice that the isomorphism, given by
$\D^\a \ga = 0$, depends on $A$.

{\it Remark 1.} We stress, in view of later discussion, that for the null connection
$A^0$, $H_0 (p_0 ) = \{ e \}$ for all points $p_0 \in P$, and thus $\Ga_0$ is the
group of covariantly constant (along $A^0$) functions from $B$ to $G$, i.e. $\Ga_0
\simeq G$. $\odot$

Given a connection $\a$ with isotropy group $\Ga_\a$ we can consider: {\tt
(i)} the {\it fixed space} of $\Ga_\a$, i.e. the space of connections
(associated with connection forms) which are left invariant by $\Ga_\a$, $$
\F (\a ) \ = \ \{ \b \in \A \ : \ \ga (\b ) = \b \ \ \forall \ga \in \Ga_\a
\} \ = \ \{ \b \ : \ \Ga_\a \sse \Ga_\b \} \ \ ; \eqno(8) $$ and {\tt (ii)}
the {\it isotropy type} of $\a$, i.e. the space of connections (associated
with connection forms) having isotropy subgroups which are $\Ga$-conjugated
to that of $\a$, $$ \S (\a ) \ = \ \{ \b \in \A \ : \ \exists \ga \in \Ga: \
\Ga_\b = \ga \Ga_\a \ga^{-1} \ \} \ \ . \eqno(9) $$

The orbit of $\a$ under $\Ga$ will be denoted by $\Ga (\a )$ (or, for ease of
notation, by $\om_\a$). If we consider the gauge transformed of $\a$, we
have easily that $\Ga_{\ga (\a )} = \ga \Ga_\a \ga^{-1}$, or in other words
$$ \Ga (\a ) \ \sse \ \S (\a ) \ \ ; \eqno(10) $$  thus, the equivalence
classes under the relation of belonging to the same isotropy type consist of
(necessarily, disjoint) unions of gauge orbits.

Actually, this equivalence relation leads to a {\it stratification} of $\A$
\ref{22} (see \ref{27-29} for more detail), pretty much as in the
classical case of compact group action on finite dimensional manifolds;
however, we have now a countable -- rather than finite -- set of strata
\ref{26-29}. 

The set $\S (\a )$ will be called the {\it stratum} of $\a$, and it can be
shown to be a smooth manifold, and actually a principal bundle \ref{29}.

\section*{2. Sobolev norms, and slices}

Let us now consider $\A$ in some more detail. We have already noticed that
it is a linear space; using the $G$-invariant scalar product in $\G$,
denoted by $\<.,.\> $, we can define a scalar product in $\A$ by $$ \( \a ,
\b \) \ = \ \int_B \[ \sum_{\mu=1}^n  \< A_\mu (x) , B_\mu (x) \>  \] \d^n x
\eqno(11) $$  where $\a = A_\mu \d x^\mu$, $\b = B_\mu \d x^\mu$. Let $|\a |
= \( \a , \a \)^{1/2}$ denote the corresponding norm.  Fixing a connection
$C^0 \in \C$ and using the induced covariant derivative $\D^0$ we define a
Sobolev scalar product and a Sobolev norm of class $k$ on $\A$ by $$ \( \a,
\b \)_k \ = \  \sum_{j=0}^k \ \[ \ \int_B \( \(\D^0 \)^j\a, \(\D^0 \)^j \b
\) d^n x \ \] \eqno(12) $$ and $$ || \a ||_k \ = \ \( \sum_{j=0}^k \int_B |
(\D^0 )^j ) \a |^2 d^n x \)^{1/ 2} \ . \eqno(13) $$  We will call $\A_k$ the
completion of $\A$ with respect to this norm. Then $\A_k$ is an Hilbert
space, and  different choices of the connection $C^0 \in \C$ give rise to
equivalent norms \ref{30-33}. If we consider also the Sobolev
completion $\Ga_k$ of the gauge group $\Ga$, we have that, for $k > k_0 =  
\[ ({\rm dim}(B) +1 )/2\]$, $\Ga_k$ is an infinite dimensional Hilbert-Lie
group modelled on a separable Hilbert space (this same condition
does also ensure that the Sobolev norm dominates the $L_\infty$ one,
see e.g. \ref{34}). 
The action of $\Ga$ on $\A$ can
be extended to a smooth action  of $\Ga_k$ on $\A_{k-1}$ and the $\Ga_k$
orbits are closed in $\A_{k-1}$ \ref{25,35,36}

From now on we will assume that all objects requiring Sobolev completion
have been completed in appropriate norms and we will write again $\A$ and
$\Ga$ instead of $\A_k$ and $\Ga_{k+1}$, and $\( \a, \b \)$ instead of $\(
\a, \b \)_k$ .

We remark that the norm (13) induce a $\Ga$-invariant distance $d$ on $\A$
defined in the usual way, i.e. as $d\( \a, \b \)  =  || \a-\b ||$ .

It is known \ref{22,23,26,27,31,36} that the action of $\Ga$ on  $\A$
admits a slice $S_\a$ at any point $\a \in \A$. The existence of a slice
guarantees the existence of a tubular neighbourhood $\U_\a$ of the $\Ga$
orbit $\om_\a$; this can be obtained by $\Ga$-transporting $S_\a$, i.e.
$\U_\a = \Ga \( S_\a \)$.

We recall that a slice at $\a$ is a submanifold $S_\a \sse \A$   such that
$\a \in S_\a$ and: \parskip=0pt\parindent=0pt

{\tt (i)} $S_\a$ is transversal and complementary to the orbit  $\om_\a$ in
$\A$ at $\a$;

{\tt (ii)} $S_\a$ is trasversal to all the $\Ga$ orbits which meet $S_\a$;

{\tt (iii)} $S_\a$ is (globally) invariant under $\Ga_\a$;

{\tt (iv)} For $\b \in S_\a$ and $\ga \in \Ga$, $\ga ( \b ) \in S_\a$
implies $ \ga \in \Ga_\a$, i.e. $\Ga_\a$ is the maximal subgroup which
leaves $S_\a$ globally invariant; this also implies $\Ga_\b \sse \Ga_\a$. 

\parskip=10pt\parindent=0pt

For later reference, we rewrite {\tt (i)} as $$ \T_\a \A = \T_\a S_\a \oplus
\T_\a \om_\a \ ; \eqno(14) $$  we also denote, again for later reference, $$
S_\a^0 \ = \ S_\a \cap \F (\a )  \ . \eqno(15) $$ 

It follows from properties {\tt (iii)} and {\tt (iv)} of $S_\a$, and the
compactness of $\Ga_\a$, that  $$ S^0_\a \ = \ \Sigma(\a) \cap S_\a \ ,
\eqno(16) $$  $$ \Sigma(\a) \cap \U_\a \ = \ \bigcup_{\b \in \om_\a } S^0_\b
\ ; \eqno(17) $$  hence, from (10),(14) and (17), we have that  $$ \T_\a
\Sigma(\a) \ = \ \T_\a \omega_\a \oplus \T_\a S^0_\a \ . \eqno(18) $$

\section*{3. Michel theorem for pure gauge theories}

Using the notation defined above, we introduce the following two definitions:

{\bf Definition 1:} {\it A gauge orbit $\om_\a$ is {\rm isolated in its
stratum} if and only if $\U_\a \cap \Sigma(\a) = \om_\a$.}

{\bf Definition 2:} {\it The $\Ga$-orbit $\om \ss \A$ is {\rm critical} if
points on $\om$ are critical for any smooth $\Ga$-invariant functional on $\A$.}

We will prove that these are actually equivalent; that is, we have the:

{\bf Theorem.} {\it A gauge orbit $\om$ is critical if and only if it is
isolated in its stratum.}

{\it Sketch of the proof.}  Let us now consider a $\Ga$-invariant functional
$\L: \A \to R$ of class $C^1$ (a special case of this is the Yang-Mills
functional), i.e. a functional such that $$ \L \( \ga (\a ) \) \ = \ \L ( \a )
\qquad \forall \ga \in \Ga \ , \ \forall \a \in \A \ ; \eqno(19) $$  its
differential at $\a$, $\d \L_\a : \T_\a \A \to R$, will be a linear and
continuous operator; as $\A$ is a Hilbert space, this will correspond to an
element $\phi_\a \in \T_\a \A \simeq \A$, such that $\d \L_\a (\b ) =
(\phi_\a , \b )$.

It is easy to see that the invariance of $\L$ implies  $$ \phi_{\ga (\a )} \
= \ \ga (\phi_\a) \ . \eqno(20) $$

This also implies that if $\ga \in \Ga_\a$, then $\d \L_\a $ is invariant
under $\ga$; that is, $$ \phi_\a \ \in \ \T_\a \F (\a ) \ \ . \eqno(21) $$

On the other side, it is clear from (19) that $(\phi_\a , \xi ) = 0$ for all
$\xi \in \T_\a \om_\a$; thus we conclude [see (14),(18)] that $$ \phi_\a \
\in \ \T_\a S_\a^0 \ \ . \eqno(22) $$

It is clear from (19) that if $\a \in \A$ is a critical point for $\L$, all
the points $\b \in \Ga (\a )$ are also critical for $\L$\footnote{$^{2}$}{It is
then natural to consider as index of $\a$ \ref{38} its index as a critical point of
the restriction of $d\L_\a$ to the slice $S_\a$, or equivalently to
$S_\a^0$.}, which justifies definition 2. The above discussion shows that if
an orbit is isolated in its stratum, then it is critical.

We could also prove the converse, i.e. that if an orbit is critical, then it
is isolated in its stratum; the proof of this would be just the same as the
one given by Michel \ref{1} for compact groups and finite dimensional
manifolds, and thus is omitted.

From the above discussion we conclude that for functionals defined on the
space of connections of a principal bundle (in Physics' language, pure gauge
theories) we have -- in the framework, with the definitions, and under the
conditions introduced so far -- the extension of Michel's theorem given
above. $\odot$

{\it Remark 2.} The simplest nontrivial case where we have nontrivial (that is, not
pure gauge) critical gauge orbits is provided by $G = SU(3)$. An analysis of strata in
this case is provided by \ref{29}; the bordering relations among strata are analyzed
in \ref{37}. $\odot$

\section*{4. Theories with matter fields}

In physical application of gauge theories \ref{39-44}, one wants to
consider not only pure gauge theories, but theories with
matter fields as well. It turns out that our main result can be
extended to this case as well, and actually that this extension does
not present any new difficulty with respect to the pure gauge case;
thus we sketch here such an extension, without repeating details
already discussed above in the pure gauge frame.

In the case of theories with matter fields, together with the
$G$-principal bundle $P$ over $B$, and the set ${\cal C}$ of connections
on $P$ (see sect.1), we should consider a vector bundle $Q$ over the same
base space $B$, having $G$ as structure group; the fiber of $Q$ will be a
vector space $E^q$ (usually ${\bf R}^q$ or ${\bf C}^q$), and the set of 
differentiable sections of $Q$ will
be denoted as $\V$. Notice this is not a complete space, and moreover that we should
impose, as for the gauge fields, a finite energy condition (these remarks will
naturally lead to consider a Sobolev space of fields in $\V$, see below).

By choosing an orthonormal frame in $E^q$, and using the reference
connection chosen in ${\cal C}$ to transport this to the fiber over any
point $x \in B$, a section $f \in \V$ is described in local
coordinates by $\phi : B \to E^q$, i.e. by $q$ fields $\phi^j (x)$, $j=1,...,q$, 
$\phi^j : B \to E$; these are usually called {\it matter fields}.

The group $G$ acts on $E^q$ by a linear representation $T$, and we
denote by $T_g $ or $T(g)$ the linear operator on $E^q$ corresponding
to $g \in G$.

We should then consider the sum bundle $P \oplus Q$ over $B$; we
denote the space of sections of this by $\Ga \oplus \V$, and a section
by $\ga \oplus f$. 

An element $\ga$ of $\Ga$ expressed in local coordinates as $g(x)$ acts
on $f \in \V$ -- expressed in local coordinates as $\phi (x) $ -- by 
$$ [\ga (f) ] (x) \ = \ T_{g(x)} \phi (x) \ . \eqno(23) $$

This means in particular that the gauge isotropy subgroup of $f$ is
$$ \Ga_f \ = \ \{Ê\ga \in \Ga \ : \ g (x) \in G_{\phi (x) } \ \ \forall x
\in B \} \ ; \eqno (24) $$
notice that this is in general a non-compact group.

We can define a natural scalar product in $\V$ using the scalar product
$\< .,. \>$ defined in $E^q$: indeed, we define
$$ (f,h) \ = \ \int_B \ \< \phi (x) , \chi (x) \> \ \d^n x \eqno(25) $$
where $f \simeq \phi (x) $, $h \simeq \chi (x) $.

We can then proceed as in section 2, and define -- using a connection
$C^0$ and the induced covariant derivative -- a Sobolev scalar product
and a Sobolev norm of class $k$ on $\V$; this $k$ will be the same as
that chosen in the analysis of connections. We will then denote by
$\V_k$ the completion of $\V$ with respect to this norm, and consider
from now on this set of sections (and drop -- as it was already done for
$\A$ -- the subscript, for ease of notation). 

Results concerning Hilbert-Lie group structures, and
existence of slices and tubular neighbourhoods in $\V$, are described e.g.
in \ref{45}.

Rather than repeating in $\V$ the analysis conducted in $\A$ -- which
would actually present some serious difficulty, as now we have
non-compact isotropy subgroups -- we will consider directly $\A \oplus
\V$. The advantage of this follows from 

{\bf Lemma.} For any $\a \oplus f \in \A \oplus \V$, $\Ga_{\a \oplus
f}$ is compact.

{\it Proof.} This is a consequence of a very simple observation, i.e.
that
$$ \Ga_{\a \oplus f} \ = \ \Ga_\a \cap \Ga_f \ . \eqno(26) $$ 
Thus, $\Ga_{\a \oplus f}$ is necessarily isomorphic to a (compact)
subgroup of the compact group $G$, since this is the case for $\Ga_\a$
(see sect.1). $\odot$

As a consequence, we can define strata in $\A \oplus \V$. We have seen
above that we can also define a Sobolev metric, and thus 
neighbourhoods in $\A \oplus \V$ are well defined.

{\it Remark 3.} It should be stressed that we cannot proceed by
defining strata in $\V$ and then ``composing'' them with strata in
$\A$; the obstacles to this -- due essentially to the non-compact
nature of $\Ga_f$ -- are discussed e.g. in \ref{19,20} (these do also provide
further detail on $\V$). $\odot$

We can then proceed as in sect.3, arriving at the same conclusions.
In particular, we define $\U_{\a \oplus f}^\eps$ as the set of sections
$\b \oplus h$ of $\A \oplus \V $ such that $|| \a - \b || + || f - h
|| < \eps$, $\om_{\a \oplus f}$ as the $\Ga$-orbit of $\a \oplus f$,
and $\S (\a \oplus f)$ as the stratum of $\a \oplus f$ in $\A \oplus
\V$. We say then that $\om_{\a \oplus f}$ is isolated in its stratum if
there is an $\eps > 0$ such that $\U_{\a \oplus f}^\eps \cap \S (\a
\oplus f) = \om_{\a \oplus f}$, and that $\om_{\a \oplus f}$ is critical
if points on $\om_{\a \oplus f}$ are critical for any $\Ga$-invariant
smooth functional defined on $\A \oplus \V$.

We have then -- as already mentiond, just following the procedure of
section 3, and thus the original proof by Michel \ref{1} -- that the
orbit $\om_{\a \oplus f}$ is critical if and only if it is isolated in
its stratum.

\section*{5. General gauge-equivariant evolution equations}

In the classical case (compact group $G$ acting on a finite dimensional
manifold $M$), it is well known that the Michel's theory and its
symmetry-based approach can be extended to consider general equivariant
dynamics rather than just variational problems; in this way one can
re-obtain\footnote{$^{3}$}{It should be stressed that these results were 
originally obtained with no use of
Michel's theory, and some years after the original Michel's paper. This seems
to mean that on the one side mathematicians were
not aware of the work of physicists on symmetry breaking, and on the other 
side also that
physicists were not able to realize the relevance of the results they knew
for other -- not so far -- fields (a partial exception being provided by
\ref{48}), i.e. for general equivariant Nonlinear Dynamics. Surely, this fact 
points out a regrettable lack of communication between
the two communities.} \ref{46,47} in particular the Equivariant
Branching Lemma \ref{48-50} (and its extension to the Hopf case) and
the Reduction Lemma \ref{51-53} (see also \ref{54}); 
most of the results in equivariant
bifurcation theory \ref{52-58} are based on these lemmata \ref{52}.

The same holds here, i.e. one could obtain the corresponding results for
equivariant vector fields on $\A$. Indeed, in this case equivariance means
that 
$$ X \( \ga (\a ) \) \ = \ \ga^* \[ X \( \a \) \] \qquad \forall \ga
\in \Ga ,~ \forall \a \in \A \ , \eqno(23) $$ 
where $\ga^*$ denote the action of $\Ga$ on $\T \A$ induced by the action of
$\Ga$ on $\A$. Thus, for $\ga \in \Ga_\a$ and $X$ an equivariant vector
field on $\A$, we have 
$$ \ga^* \[ X(\a ) \] = X \( \ga ( \a ) \) = X (\a )
\ . \eqno(24) $$ 
We have thus proven the lemma below:

{\bf Lemma.} {\it Let $X : \A \to \T \A$ be a vector field  on $\A$,
equivariant under the gauge group $\Ga$; then, $X \( \a \) \in \T_\a \F (\a
)$.}

From this lemma one could obtain immediately ``infinite dimensional
versions'' of the Equivariant Branching Lemma and of the Reduction Lemma, in
essentially the same way as in the finite dimensional case: indeed, the
relevant feature here is the invariance under an equivariant flow (which
includes the gradient flow for an invariant potential or functional) of the
closure of a subspace defined by invariance properties (the strata or even
the spaces $\F (\a )$ considered above). 
As $\F (\a )$ is in the closure of $\S (\a )$, we also have immediately the

{\bf Corollary.} {\it Let $X : \A \to \T \A$ be a vector field  on $\A$,
equivariant under the gauge group $\Ga$;  then, $X \( \a \) \in \T_\a \S (\a
)$.}

{\it Remark 4.} It should be stressed that the limitation to gauge theories
and gauge invariant functionals (or, in this last lemma, gauge equivariant
vector fields)  was only dictated by the physical interest of this case; it
turns out that one could as well extend Michel theory to the infinite
dimensional setting (under suitable technical conditions) irrespective of
the gauge structure. In this way, and using the symmetry theory for
differential equations \ref{59-63}, one could e.g. identify
functions which are solutions for all the differential equations having a
given symmetry (in a precise sense, amounting to commuting flows in a
generalized function space) \ref{64}. $\odot$

{\it Remark 5.} It is maybe also appropriate to stress, in this respect,
that although the results based on $\F (\a )$ are, in principle, stronger
than those based on $\S (\a )$ (because the partition into sets $\F (\a )$ is
finer\footnote{$^{4}$}{More precisely, to speak about a finer partition we should
consider the closures of strata, since -- as already remarked -- with our definition
the $\F (\a ) $ belong to the closure of $\S (\a )$, not necessarily to $\S (\a )$
itself.} than the one into strata $\S (\a )$), from the physical point of view
it is preferable, and more natural, to work with strata: on the one side,
this mantains the identification among gauge-equivalent objects; and on the
other side this permits to pass to $\A / \Ga$ (also called the {\it configuration
space} in physical literature), which
is natural for a number of physical considerations.

On the other side, from a mathematical point of view one can be well
justified in adopting the $\F (\a )$ point of view, as this is in general
more powerful. This is quite similar to the situation in the finite
dimensional case: from the physical point of view, the Michel approach is
``the'' natural one, as the orbit space $\Omega = M/G$ is the interesting
one, and the approach based on the $\F (\a )$ cannot be applied to $\Omega$ if not
passing through consideration of the strata. However, in other frames, 
e.g. in bifurcation theory \ref{52,53} and anyway
when one is interested in $M$ and not so much in $\Omega$, the stronger
reduction provided by the invariance of $\F (\a )$ proves very useful as
it allows a reduction to smaller dimensional submanifolds.  $\odot$

\vfill\eject
\section*{6. Reduction and symmetry breaking.}

In physical applications of gauge theories, the theory -- i.e. the functional $\L$ to be
extremized -- can depend on a control parameter, and one is specially
interested in the occurrence of spontaneous {\it symmetry breaking}, i.e. in
the case where $A^0$ is a critical point for all value of the control parameter $\la$, 
and a minimum for $\la < \la_0$, but for $\la > \la_0$ it is {\it not} a minimum and
is thus unstable; minima are then realized by other, symmetry-breaking,
critical points; we call these $A^* (\la )$ and denote the connection forms by $\a
(\la )$.

We will consider a second order phase transition, i.e. the case where we will have 
$$ \lim_{\la \to \la_0^+} A^* (\la ) = A^0 \ . \eqno(25) $$
In mathematical language, this corresponds to a {\it bifurcation}, and we have a
continuous branch of bifurcating solutions.
 
Typically, these $A^* (\la )$ ($\la > \la_0$) have a symmetry which is less 
than the one of $A^0$ (this is invariant under any constant section, see 
remark 1 above), but which is the
same for a whole set of values of $\la$: thus, $A^* (\la )$ for $\la \in (\la_0 ,
\la_1 )$ belong to the same stratum. This also means that the $A^* (\la )$
correspond to gauge orbits which cannot be isolated in their stratum, so that the
gauged version of Michel theorem given above cannot be used.

However, we can still be able to obtain relevant informations by
``quotienting out'' the unavoidable -- and thus, in a sense, trivial --
degeneracy of $\Ga_{\a (\la )}$ in the direction ``along the bifurcating branch''. 
The idea will be essentially to separate the variation in
this direction (on which symmetry considerations can be of no use) and
the one in the other, transversal, directions\footnote{$^{5}$}{Notice that the
application of Michel theorem to $G = SU(3)$ met the same problem; in that
case, the direction ``along the branch'' would simply be the radial one, and 
thus one would simply consider the unit sphere in $\G = su(3)$.}. 

Thus we suppose that there is a stratum $\mu (\la )$ (notice this can depend on
$\la$) in $\A$ such that $A^0$ belongs
to the border of $\mu (\la )$ for all $\la \in (\la_0 , \la_1 )$, see above, 
and that $\mu (\la ) / \Ga $ is a one-dimensional manifold (the latter one is a
strong assumption, which presumably could be somewhat relaxed). 

We are then guaranteed of the vanishing of $(\de \L ) [\a ]$ in directions transversal
to $\mu (\la )$; this means that in this case 
we can split the search for critical points of the
$\Ga$-invariant functional $\L$ in two steps: {\it (i)} we determine
strata $\mu (\la )$ such that $\mu (\la ) / \Ga $ is one-dimensional; 
{\it (ii)} once we have determined such a $\mu (\la )$, 
we consider the restriction of $\L$ to $\mu (\la )$, and actually to 
$\mu (\la ) / \Ga $: i.e. we
reduce to a variational problem in one dimension. 

Notice that when the functional is confining (i.e. we are guaranteed of the
existence of a ball ${\cal B}$ in $\A$ such that $\L > \L [A^0 ]$ on $\A \backslash
\B$ and $\delta \L$ is inward on
$\pa {\cal B}$), we are guaranteed of the existence of a critical point in
$\S (\a )$, and when we add the condition that $A^0$ is unstable for $\la > \la_0$, 
we are actually guaranteed of the existence of critical gauge orbits in $\mu (\la )$.

This approach can be seen as an application of the ``Symmetric Criticality
Principle'' of Palais \ref{65,66}; or also -- more simply -- as 
a generalization of the equivariant branching
lemma \ref{48-50}, well known in equivariant bifurcation theory 
\ref{52-58}.

{\it Remark 6.} The symmetric criticality principle 
can also be applied -- and is actually originally
formulated \ref{65} -- in what we called the $\F (\a )$ frame. In this
way, it becomes even more powerful for what concerns general
gauge-equivariant evolution problems. However, the same considerations
presented in remark 5 apply here, i.e. in the context of physical gauge
theories one should identify gauge-related spaces and thus work with strata
rather than with fixed spaces. $\odot$

\section*{7. Extensions, generalizations, and discussion.}

We would like to conclude this note by a series of remarks concerning
possible extensions of our results, comparison of these with existing
results on critical points of gauge functionals, and/or comparison with the
Michel theory for the classical case (we recall by this we mean the
finite dimensional case, i.e. a compact Lie group $G$ acting on a finite
dimensional smooth manifold $M$).

{\it Remark 7.} First of all, we notice that in the classical case, one can
apply the classical tools of variational analysis (e.g.
Lyusternik-Shnirelman and Morse theories; see e.g. \ref{67} and refernces
therein, and \ref{68}), 
and also stratified Morse theory
\ref{69} to obtain further information on the number  and nature of
critical points in (the closure of) each stratum; see e.g. \ref{70} for a
simple example, and especially \ref{71} for a recent --  physically
relevant and mathematically more interesting -- application of  this
approach. It appears that the same holds -- with increased technical
difficulties  to be expected, as we are now in an infinite dimensional
setting -- for our present extension. $\odot$

{\it Remark 8.} We would also like to mention that 
in the classical case one is able, modulo some
further assumption on the $G$-action on $M$ -- to project an equivariant
vector field onto a field in the orbit space\footnote{$^{6}$}{Notice that in
general $\Omega$ is not a manifold, but it is a stratified manifold, i.e.
the union of strata in the geometric (Whitney) sense,  which themselves are
manifolds. Notice also that our theorem guarantee that the projected field
will be tangent to these strata, provided one proves the 
Whitney strata are union of Michel strata (see \ref{46,47}).} $\Omega = M/G$. 
In the present setting, it is known \ref{22,29} that $\A$ can be
decomposed as the union of principal $G_j$-bundles (the $G_j$ being certain
subgroups of $G$); moreover, we know that $\A$ has a proper isotropy
stratification. Thus it would be worth studying if there are reasonable
conditions under which any equivariant vector field $\phi^* : \A \to \T \A$
can be projected to some vector field $\psi^* : \A / \Ga \to \T \( \A / \Ga
\)$ (wherever this makes sense). $\odot$

{\it Remark 9.} In this note we adopted the point of view of studying the singular
strata (i.e. orbits which are not in the generic stratum, or equivalently reducible
connections), arguing that -- as it is shown by our extension of Michel theorem --
these will provide ``generic'' solutions in the presence of symmetry, precisely
because they are singular, or even better ``as singular as possible''.

A different approach is also possible, using again the stratification but proceeding
the other way round, i.e. considering first the generic orbits (i.e. $\Ga_A \simeq
G$), then connections which are reducible but with large $\Ga_A$, and so on, and at
each stage restricting attention to these ``minimally reducible'' connections, on
which one can, by the procedure very shortly described in a moment, apply the results
holding for reducible connections \ref{22,23} 
provided the theory is set on a reduced bundle
\ref{29}. Let us consider a reducible connection $\a$; 
from (5), the $\Ga_\a$ would depend on the
reference point 
$$ p_0 = \( x_0 , \ga (x_0 ) \) \in \pi^{-1} (x_0 ) \ss P \ ; $$
choosing a different reference point $g p_0$ on the same fiber will
change $H_A (p_0 ) $ to $H_A (g p_0 ) = g H_A (p_0 ) g^{-1} $, and it
follows easily that $C_G [H_A (g p_0 ) ] = g \( C_G [H_A (p_0 ) ] \)
g^{-1} $. Thus, nothing changes -- in $H_A$ and hence in $\Ga_\a$ -- when
we act by $$ g \in J_A \ \equiv \ C_G [ C_G [H_A (p_0 ) ] ] \ . $$ 
We can then consider a subbundle $P_A$ over the same base
space $B$ and fiber $\pi^{-1} (x) \simeq J_A \sse G$; the connection
$\a$ is irreducible over this subbundle. We can thus analyze more and more 
singular strata in $\A$ by proceding along chains of subbundles \ref{29}; 
strata can be analyzed in terms of these, and bordering relations can be expressed as
relations among characteristic classes \ref{37}. This approach could also
be extended to gauge theories with matter fields. $\odot$

{\it Remark 10.} At the same time as the Michel-Radicati approach \ref{1-4}, a
different approach, also based on geometry of group action, was proposed by Cabibbo
and Maiani \ref{72}; this is not equivalent to the Michel-Radicati approach, but
seems to somewhat be a precursor of the symmetric criticality principle of Palais 
\ref{65} mentioned above. $\odot$

{\it Remark 11.} In the present note, we have only mentioned those results on
the geometry of the gauge 
orbit space (configuration space) $\A / \Ga$ that we needed; however, the geometry of
this space has been studied quite in detail. We refer to
\ref{22-30,35,37}, already
mentioned, and e.g. to \ref{73-77} for further detail 
and bibliographic indications. $\odot$

{\it Remark 12.} It should be stressed (see also the remarks below) that the
results presented here -- and the method they provide to search for critical points of
gauge-invariant functionals -- are settled in general terms, i.e. for general (smooth)
gauge-invariant functionals; this is in principles an advantage of them.
However, this fact can also
be considered as a weakness of our approach: that is, in this way we fail to take
advantage of the specific features of the functionals of interest in physical gauge
theories (see the appendices). $\odot$

{\it Remark 13.} We would like to stress that the approach
presented here is quite general, and can be applied to any base manifold $B$
and Lie group $G$; similarly, it can be used to look for any kind of
critical point, and not only for minima and/or for selfdual or anti-selfdual
critical points. 

In this respect, we recall that the critical points
which are not minima can be quite relevant physically: this is the case e.g.
in the WKB theory (see also the Duistermaat-Heckman localization theory
\ref{78-80},  guaranteeing that in certain cases the WKB approximation is exact).

Similarly, we recall that the celebrated results of
Bourguignon and Lawson \ref{31-33}, stating that all weakly stable
critical points of Yang-Mills functionals are self-dual or anti-self-dual, apply under
precise conditions (which enclose the physically relevant cases); in
particular, they only hold in dimension four, and for $G=SU(2)$ and $SU(3)$
[with extensions to $U(3)$ and $SO(4)$], and do not deal with unstable
critical points. 

Moreover, we also mention that the present approach could also be
useful (as it localizes the critical points in strata and thus provides a
natural parametrization of them) in determining the moduli space for
instanton/monopole solutions, and the Atiyah-Hitchin metric.  $\odot$

{\it Remark 14.} As already mentioned in the introduction, previous attempts to extend
Michel theory to the gauge case exist, and were able to provide some results
\ref{19,20}; however, they suffered from severe limitations. Essentially, these
only considered functionals on $\V$ (moreover, overlooking any contact structure
\ref{59,62,75} when considering dependence on derivatives). These limitations
followed from tackling the problem by the matter field sector (see the remarks in
sect.4 for the difficulties with this approach); in the present complete -- and
geometrical -- setting, it is the strong geometrical structure present in $\A$, i.e.
in the pure gauge sector, which allows to tame the problems arising in $\V$ and which
forced the previous attempts \ref{19,20} to such a limited scope. $\odot$

{\it Remark 15.} It should also be stressed that the results given here
extend to the case of a gauge functional depending on higher order
(covariant) derivatives, thus extending substantially the case of proper
Yang-Mills functionals. $\odot$

{\it Remark 16.} More in general, apart from technical problems, the whole
construction considered here, and by Michel theory, relies on the concept of
group action on a manifold, and of invariant functionals (or equivariant
ones once we consider e.g. the gradients), together with a suitable topology
and an appropriate ``agreement'' between the manifold and group topologies
(the requirement that group orbits are proper submanifolds of the manifold);
it is quite clear that this is a very general situation, and thus that the
reach of Michel theory is correspondingly ample. $\odot$

Thus, in conclusion, we believe Michel theory provides a new approach to a
number of questions  which are extremely relevant for Physics, and possibly
also for Mathematics;  needless to say, the credit for it should go
to the original geometric view it embodies \ref{1}, and not to the simple,
albeit useful, extension given here.

\vfill

\section*{Acknowledgements} 

{\petitrm \parindent=10pt \parskip=0pt
This work was started at the Mathematische Forschunginstitut in
Oberwolfach; our staying in M.F.O. was supported by the Volkswagen Stiftung under the
``Research in Pairs'' program. We would like to thank the Director of M.F.O., prof. M.
Kreck, for hospitality.

The work was completed during the stay of one of us (G.G.) in I.H.E.S.; 
warm thanks go to the Director, prof. J.P. Bourguignon, and all the personnel of the
Institute for the invitation and hospitality.

The work of P.M. received essential support from GNFM-CNR through travel funds to
Oberwolfach, Loughborough and Bures sur Yvette. We also benefited from travel grants
by Loughborough University.

In the preparation of this work we have bothered several friends and colleagues with
our questions; we would like in particular to thank C. Bachas and A. Chakrabarti.}

\vfill\eject

\section*{Appendix A. Special solutions in theories with matter fields.}

In this note, we considered theories defined by any smooth gauge-invariant functional
on $\A$ or on $\A \oplus \V$; however, in physical gauge theories -- i.e. Yang-Mills
theories \ref{39-44} -- one mostly meets functionals with a precise form.

In this and the following appendices we want to show how our results apply to proper
Yang-Mills theories, in order to get special solutions of interest, and recover the
Michel-Radicati results \ref{3,4} in the SU(3) case. In this first
appendix we want to show how we can by our result get some special, and
relevant, solutions in the case of theories with matter fields, once we have
determined critical points of the pure gauge part of the theory.

Thus we will consider some vector space $E$ and a vector bundle over $B$ with fiber
$E$ having $G$ as structure group. 
When we include matter fields $\phi : B \to E$  in the theory, 
we have -- restricting to the 
physically relevant situation -- a functional 
$$ \L \ = \ \L_g (A,\phi ) + \L_p (\phi ) + \L_{ym} (A) \ , \eqno(A1) $$
where the three parts of the functional are defined as
$$ \eqalign{ \L_g = &   \int_B | (\grad_A \phi ) (x) |^2 \, \d^n x \cr
 \L_p = &  \int_B V[\phi (x) ] \, \d^n x \cr
 \L_{ym} = & \int_B | F_{\mu \nu} [A(x)] |^2 \, \d^n x \cr}  \eqno(A2) $$ 
with $V$ a smooth function (potential) $V: E^q \to R$, and 
the norms defined via the appropriate scalar products. We will consider
$B = R^n$, and $A^0$ is the flat connection.

In looking for critical points of this functional, we can look for a special class of
these, i.e. we look for functions $\phi
(x) $ and $A_\mu (x)$ which extremize the three parts separately: these will then be a
satisfactory extremizing solution for $\L$ (although not all the solutions will be
obtained in this way).

We can start from $\L_{ym}$, and suppose we have determined a connection $\a$ (i.e.
the functions $A_\mu (x)$, valued in $\G$) which is critical for $\L_{ym}$; 
in particular, these could be determined by means of our previous results.

Let us now consider $\L_p$; we consider $\F$, the set of sections (in suitable Sobolev
class) of the bundle over $B$ having $E$ as fiber, and $M \ss E$ the set of points
on the fiber on which $V$ has extremal points. We then define $\F_M \ss \F$ as the 
set of sections $\phi$ such that $\phi (x) \in M$ for all $x \in B$. 
It is clear that any $\phi \in \F_M$ is critical for $\L_p$.

We should now consider $\L_g$, the only part in which $A$ and $\phi$ interact: this
will be extremal -- actually minimal -- if $\phi$ is covariantly constant along $\a$.
We will try to build a $\psi$ which satisfy this condition and is in $\F_M$.

Thus, we consider a reference point $x_0 \in B$, and choose $\psi (x_0 ) = m_0 \in M$;
then we define $\psi (x)$ by the condition $\grad_A \psi = 0$, i.e. as the solution to
the equation
$$ {\pa \psi \over \pa x^\mu } \ = \ - A_\mu (x) \, \psi (x) \eqno(A3) $$
To this end, we choose for any $x \in B$ a parametrized path $\xi (s)$ such that
$\xi (0) = x_0$ and $\xi (1) = x$. Now, the required solution is provided by
$$ \psi (x) \ = \ g(x) m_0 \eqno(A4) $$
where $g(x)$ is defined as
$$ g(x) \ = \ - \int_0^1 A_\mu [ \xi (s) ] \, {\pa \xi^\mu \over \pa s} \, \d s 
\ . \eqno(A5) $$

Clearly, this depends on the choice of the path $\xi (s)$: that is, such $g(x)$ is
defined up to an element of the holonomy group of $\a$ at the point $p=(x,g(x) )$;
this was previously noted as $H_A (p)$. Correspondingly, $\psi (x)$ is defined up to
the action of $H_A (p) $ on $g(x)$: thus $\psi (x)$ is well defined only if at any
point 
$$ H_A [ x , g(x) ] \sse G_{\psi (x) } \ . \eqno(A6) $$

By general results on $H_A$ along covariantly constant curves \ref{82}, this
condition can be checked by looking at the fiber over $x_0$ alone; moreover, we have
\ref{22} that
$$ H_A (x_0 , g ) \ = \ g H_A (x_0 , e ) g^{-1} \eqno(A7) $$
and thus we can just require that
$$ H_A ( x_0 , e ) \ss G_{m_0} \ . \eqno(A8) $$

It should be recalled \ref{22} that a close relationship exists between the isotropy
group $\Ga_\a$ and the holonomy group $H_A$: indeed,
$$ \Ga_\a \ = \ \{ \ga \in \Ga \ : \ \D_A (\ga ) = 0 \ ; \ \ga (x_0 ) \in C_G \[ H_A
(p_0) \] \} \eqno(A9) $$
where $C_G (H)$ is the centralizer of $H \sse G$ in $G$, and $p_0 = (x_0 , g)$ a
reference point in $\pi^{-1} (x_0 ) \ss P$.

In this way we have determined, given $A$, which extremal points 
$m \ss M$ are suitable for the
construction described above.

\vfill\eject

\section*{Appendix B. Solutions associated to the null connection}

Let us further restrict the setting of appendix A; that is, consider the critical
orbit in $\A$ corresponding to pure gauges, or in other words the gauge orbit $\om_0
= \Ga (A^0 )$. Notice that in order to be sure $\om_0$ is isolated in its stratum, and
thus critical, it suffices that the action of $\Ga$ is free (so that only pure gauges
have $\Ga_\a \simeq G$), as it is usually the case in physical applications.

In this case, we have immediately
$$ H_A (x_0,g) = \{ e \} ~~~~~ \forall g \in G \eqno(B1) $$
$$ \Ga_0 = \{ \ga \in \Ga \ : \ \D^0 (\ga ) = 0 \} \simeq G \eqno(B2) $$
and also the $f$ given by the construction in appendix A will be covariantly constant
along $\D^0$, i.e. $\phi^j (x) = {\rm const}$ for any $j$ and on any chart on $B$.

Thus, for the corresponding matter fields, we know that we can restrict to consider 
$$ \V_0 \ = \ \{ f \in \V \ : \ \D^0 (f) = 0 \} \eqno(B3) $$
or, in other words, we can simply consider $f (x_0 ) = \Phi$; notice that $\Ga_0$ acts
naturally on $\V_0$ and, with obvious notation, 
$$ \[ \ga (f) \] \, (x_0) \ = \ g_0 \, \Phi \ . \eqno(B4) $$

We can thus apply the classical version of Michel theorem in order to determine the $f
\in \V_0$ which are isolated in their stratum, and thus the results of
\ref{1-4} immediately apply to this case. 

Notice that, to be completely rigorous, we should remove the radial
degeneracy (indeed $f$ and $\la f$, with $\la \in R$, will have the same isotropy group)
or equivalently consider ``critical directions'' (see sect.6). 

\vfill\eject
\section*{Appendix C. The SU(3) case}

The motivation for Michel's theory was provided, as already mentioned in
the introduction, by the $SU(3)$ theory of hadronic interactions
\ref{3,4,7}. In this case, the relevant group action is the adjoint
representation of $SU(3)$; this acts on the space $\M$ of three-dimensional
unitary traceless matrices, and thus on $R^8$: an invariant potential is
then a function $V : R^8 \to R$ such that $ V (g\cdot x) = V(x)$, the action
of $g \in SU(3)$ on the matrix $M$ being given by $g : M \in R_g M
R_g^{-1}$, where $R_g$ is the matrix representing $g$ in the standard
three-dimensional representation of $SU(3)$. In this case $\Omega = \M / G$
is two dimensional.

One considers then the restriction of $V$ to the unit sphere $\M_0$ in $\M$
under the appropriate scalar product, i.e. the $G$-invariant scalar product in the
algebra $su(3)$; this is given by $(A\cdot B) = (1/2) \, {\rm Tr} (A.B)$. One
can then check \ref{3,4,53} that the points in $\M_0 / G$
corresponding to physical particles (the $SU(3)$ octet) are isolated in
their strata, and thus the associated $G$-orbits identify critical orbits
for any invariant potential on $\M_0$, and hence directions of symmetry
breaking for any invariant potential on $\M$. This shows that predicting
the actual directions of symmetry breaking in hadronic interactions is not
a virtue of any particular model, but only of its invariance properties
\ref{5,7}.

We will now discuss in some detail the extension of this result to the
complete gauge setting considered here; such an extension is actually
immediate (so much that Michel and Radicati did not feel any need to
discuss it) but we believe it can help in fixing the idea about our present
result to see how this applies in a very well known case.

Let us consider the structure of SU(3) subgroups; this is given by the following
diagram:
$$ SU(3) \to SU(2) \times U(1) \matrix{\nearrow\cr\searrow\cr} \matrix{SU(2) \cr ~ \cr
 U(1) \times U(1) \cr} \matrix{\searrow \cr \nearrow \cr} U(1)
\to \{ e \} \ . \eqno(C1) $$

In the spirit of studying symmetry breaking from the trivial solution $(\a , f ) =
(A^0 , 0 )$ -- and having already considered the case where symmetry is broken only in
the matter sector, while that where it is broken only in the gauge sector is
equivalent to the pure gauge case -- we focus on $G_0 = SU(2) \times U(1) $; thus, we
consider critical gauge orbits in $\A$ with 
$$ \Ga_A \ = \ SU(2) \times U(1) \ \equiv \ G_0  \ . \eqno(C2) $$
For such a critical orbit -- i.e. fixing a critical $A$ which satisfies (C2) -- we have
$$ H_A (p_0 ) = Z (\Ga_A ) = U(1) \times U(1) = Z(G) \ . \eqno(C3) $$

From the requirement that $H_A \sse G_{m_0}$ (i.e. in this case $H_A \sse G_0$), 
see (A9), we immediately reduce to three
possibilities for $\Ga^{(A)}_f = \{ \ga \in \Ga_A \ : \ \ga (f) = 0 \}$; i.e., this
can be isomorphic to either $G$, either $G_0$, either $Z(G)$. In the first case we
just have gauge fields (no symmetry breaking in the matter sector), while in the third
case we would need further hypotheses to be sure the critical point located in the
closure of the stratum is not actually on its border (and thus in the most singular
stratum with isotropy type $G_0$). Thus, the case of interest is that of 
$$ \Ga^{(A)}_f \ = \  \Ga_{(\a , f)} \ = \ SU(2) \times U(1) \ . \eqno(C4) $$

The construction of appendix A tells how to build a section $f \in \V$ which is
critical for (A1) in this case; more generally, define
$$ \V_A \ = \ \{ f \in \V \ : \ \Ga_A \cap \Ga_f = \Ga_A \} \ = \ \{ f \in \V \ : \
\Ga_A \sse \Ga_f \}  \eqno(C5) $$
(this is analogous to the $\V_0$ considered above); we consider then, for $\om_\a$
critical in $\A$, the restriction $\L_A$ of $\L : \A \oplus \V$ to 
$\a \oplus \V_A \simeq \V_A$; by the
Symmetric Criticality Principle \ref{65,66}, the critical points of $\L_A$ will
also be critical points of $\L$.

Within $\V_A$, we can proceed as for  $\V_0$ in appendix B, and thus use again the
classical version of Michel theorem; once again, this is due to the fact that $\Ga_A$
is canonically isomorphic to a compact subgroup of $G$.

Thus we conclude that for SU(3) (acting on $E^q$ by a free representation $T$) we
always have, in the presence of symmetry breaking in both the pure gauge and the
matter sectors, critical orbits $\om_{\a , f}$ with the symmetry of both $\a$ and $f$
given by a group isomorphic to $SU(2) \times U(1)$; these correspond again to the
critical directions determined by Michel and Radicati \ref{1-4}.

\vfill\eject

\section*{References} 

\parindent=10pt
\parskip=0pt

\bigskip\bigskip
\def\CMP{{\petitit Comm. Math. Phys. }}
\def\ref#1{\item{#1.} }
\font\petitrm =  cmr8 
\font\petitit = cmsl8
\font\petitbf = cmbx8
\baselineskip=10pt

\def\tit#1{{\petitrm ``#1''}}
\def\tit#1{{``#1''}}
\petitrm

\ref{1} L. Michel, \tit{Points critiques de fonctions invariantes sur
une G-vari\'et\'e}, {\petitit Comptes Rendus Acad. Sci. Paris} {\petitbf
272-A} (1971), 433-436

\ref{2} L. Michel and L. Radicati, \tit{Breaking of the $SU_3 \times SU_3$ symmetry
in hadronic Physics};  in M. Conversi (ed.), \tit{Evolution of particle Physics
(E.Amaldi Festschrift)}, pp. 191-203, Academic Press 1970

\ref{3} L. Michel and L. Radicati, \tit{Properties of the breaking of
hadronic internal symmetry}, {\petitit Ann. Phys. (N.Y.)} {\petitbf 66}
(1971), 758-783

\ref{4} L. Michel and L. Radicati, \tit{The geometry of the octet},
{\petitit Ann. I.H.P.} {\petitbf 18} (1973), 185

\ref{5} L. Michel, \tit{Nonlinear group action. Smooth action of compact Lie groups
on manifolds}, in ``Statistical Mechanics and Field Theory'', R.N. Sen and C. Weil
eds., Israel University Press, Jerusalem 1971

\ref{6} L. Michel, \tit{Les brisure spontan\'ees de sym\'etrie en
physique}, {\petitit J. Phys. (Paris)} {\petitbf 36} (1975), C7 41

\ref{7} L. Michel, \tit{Symmetry defects and broken symmetry.
Configurations. Hidden symmetry}, {\petitit Rev. Mod. Phys.} {\petitbf 52}
(1980), 617-651

\ref{8} M. Abud and G. Sartori,  \tit{The geometry of
orbit space and natural minima of Higgs potentials}, {\petitit Phys. Lett. B}
{\petitbf 104} (1981), 147-152

\ref{9} M. Abud and G. Sartori, \tit{The geometry of spontaneous symmetry
breaking}, {\petitit Ann. Phys.} {\petitbf 150} (1983), 307-372

\ref{10} G. Sartori and V. Talamini, \tit{Universality in orbit spaces of compact
linear groups}, \CMP {\petitbf 139} (1991), 559-588

\ref{11} G. Sartori and G. Valente, \tit{Orbit spaces of reflection groups with 2,3
and 4 basic polynomial invariants}, {\petitit J. Phys. A} {\petitbf 29} (1996), 193-223

\ref{12} G. Sartori, \tit{Geometric invariant theory. A model-independent approach to
spontaneous symmetry and/or supersymmetry breaking}, {\petitit Riv. N. Cim.} {\petitbf
14} (1991) no. 11, 1-120

\ref{13} M.J. Field and R.W. Richardson, 
\tit{Symmetry-breaking and the maximal isotropy
subgroup conjecture for reflection groups}, 
{\petitit Arch. Rat. Mech. Anal.} {\petitbf 105}
(1989), 61-94

\ref{14} M.J. Field and R.W. Richardson, 
\tit{Symmetry-breaking in equivariant bifurcation
problems}, 
{\petitit Bull. A.M.S.} {\petitbf 22} (1990), 79-84

\ref{15} M.J. Field and R.W. Richardson, 
\tit{Symmetry-breaking and branching patterns in
equivariant bifurcation theory -- I}, {\petitit Arch. Rat. Mech. Anal.} {\petitbf 118}
(1992), 297-348

\ref{16} M.J. Field and R.W. Richardson, 
\tit{Symmetry-breaking and branching patterns in
equivariant bifurcation theory -- II}, {\petitit Arch. Rat. Mech. Anal.} {\petitbf 120}
(1992), 147-190

\ref{17} G. Bredon, {\petitit Compact transformation groups}, Academic Press, 1972

\ref{18} M.J. Field, {\petitit Symmetry breaking for compact Lie groups}, 
{\petitit Mem. A.M.S.} {\petitbf 120} n.574 (1996), 1-170

\ref{19} G. Gaeta, \tit{Michel's theorem and critical section
of gauge functionals}, {\petitit Helv. Phys. Acta} {\petitbf 65} (1992),
922-964

\ref{20} G. Gaeta, \tit{Critical sections of gauge functionals: a symmetry
approach}; {\petitit Lett. Math. Phys.} {\petitbf 28} (1993), 1-11

\ref{21} G. Gaeta, \tit{Equivariant branching lemma: dynamical systems, evolution
PDEs, and gauge theories}, {\petitit Acta Appl. Math.} {\petitbf 28} (1992), 43-68

\ref{22} I.M. Singer, \tit{Some remarks on the Gribov ambiguity}, \CMP
{\petitbf 60} (1978), 7-12

\ref{23} I.M. Singer, \tit{The geometry of the oprbit space for non-abelian gauge
theories} {\petitit Physica Scripta} {\petitbf 24}
(1981), 817-820

\ref{24} M.S. Narashiman and T.R. Ramadas, \tit{Geometry of SU(2) gauge fields} 
\CMP {\petitbf 67} (1979), 121-136

\ref{25} G. Dell'Antonio and D. Zwanziger, \tit{Every gauge orbit passes
inside the Gribov horizon}, \CMP {\petitbf 138} (1981), 291-299

\ref{26} J.P. Bourguignon, \tit{Une stratification de l'espace des
structures riemanniennes}, {\petitit Comp. Math.} {\petitbf 30} (1975), 1-41

\ref{27}  W. Kondracki and P. Sadowski, \tit{Geometric
structure on the orbit space of gauge connections}, {\petitit J. Geom. Phys.}
{\petitbf 3} (1986), 421-434

\ref{28} W. Kondracki and J.S. Rogulski, \tit{On the
stratificatrion of orbit space for the action of automorphisms on connections},
{\petitit Diss. Math.} {\petitbf 250} (1986), 1-62 

\ref{29} A. Heil, A. Kersch, N. Papadopolous, B. Reifenhauser and
F. Scheck, \tit{Structure of the space of reducible connexions for
Yang-Mills theories},  {\petitit J. Geom. Phys.} {\petitbf 7} (1990),
489-505

\ref{30} K.B. Marathe and G. Martucci, \tit{The geometry of gauge fields},
{\petitit J. Geom. Phys.} {\petitbf 6} (1989), 1-106

\ref{31} H.B. Lawson , {\petitit The theory of gauge fields in four
dimensions}, A.M.S. (Providence) 1985

\ref{32} J.P. Bourguignon and H.B. Lawson, \tit{Stability and isolation
phenomena for Yang-Mills fields}, \CMP {\petitbf 79} (1981), 189-230

\ref{33} J.P. Bourguignon, H.B. Lawson and J. Simons, \tit{Stability and
gap phenomena for Yang-Mills fields}, {\petitit Proc. Natl. Acad. Sci. USA}
{\petitbf 76} (1979), 1550-1553

\ref{34} H. Brezis, {\petitit Analyse fonctionelle}, Masson, Paris 1983

\ref{35} P.K. Mitter and C.M. Viallet, \tit{On the bundle of connections and
the gauge orbit manifold in Yang-Mills theory}, \CMP {\petitbf 79} (1981),
457-472

\ref{36} J. Sniatycki, G. Schwarz and L. Bates, \tit{Yang-Mills and Dirac
fields in a bag, constraints and reduction}, \CMP {\petitbf 176} (1996),
95-115

\ref{37} C.J. Isham, \tit{Space-time topology and spontaneous symmetry breaking},
{\petitit J. Phys. A.} {\petitbf 14} (1981), 2943-2956

\ref{38} M.F. Atiyah and J.D.S. Jones, 
\tit{Topological aspects of Yang-Mills theory}, \CMP
{\petitbf 61} (1978), 97-118

\ref{39} C.N. Yang and R.L. Mills, \tit{Conservation of isotopic spin and isotopic
gauge invariance}, {\petitit Phys. Rev.} {\petitbf 96} (1954), 191-195

\ref{40} E.S. Abers and B.W. Lee, \tit{Gauge theories}, {\petitit Phys. Rep.} 
{\petitbf 9}
(1973), 1-141

\ref{41} T. Eguchi, P.B. Gilkey and A.J. Hanson, \tit{Gravitation, gauge theories,
and differential geometry}, {\petitit Phys. Rep.}
{\petitbf 66} (1980), 213-393

\ref{42} W. Drechsler and M.E. Mayer, \tit{Fibre bundle techniques in gauge
theories}, {\petitit Lect. Notes Phys.} {\petitbf 67}, Springer 1977

\ref{43} M. Daniel and C.M. Viallet, \tit{The geometrical setting of gauge theories
of Yang-Mills type}, {\petitit Rev. Mod. Phys.} {\petitbf 52} (1980), 175-197

\ref{44} C.J. Isham, {\petitit Modern differential geometry for physicists}, World
Scientific 1989

\ref{45} R.S. Palais and C.L. Terng, {\petitit Critical point theory and submanifold
geometry}, {\petitit Lect. Notes Math.} {\petitbf 1353}, Springer 1988

\ref{46} G. Gaeta, \tit{A splitting lemma for equivariant dynamics},
{\petitit Lett. Math. Phys.} {\petitbf 33} (1995), 313-320; 

\ref{47} G. Gaeta, \tit{Splitting equivariant dynamics}, {\petitit Nuovo
Cimento B} {\petitbf 110} (1995), 1213-1226

\ref{48} G. Cicogna, \tit{Symmetry breakdown from bifurcation}, {\petitit
Lett. Nuovo Cimento} {\petitbf 31} (1981), 600-602

\ref{49} A. Vanderbauwhede, {\petitit Local
bifurcation and symmetry}, Pitman (Boston) 1982

\ref{50} G. Cicogna, \tit{A nonlinear version of the equivariant bifurcation lemma},
{\petitit J. Phys. A} {\petitbf 23} (1990), L1339-L1343

\ref{51} M. Golubitsky and I.N. Stewart, \tit{Hopf bifurcation in the
presence of symmetry}, {\petitit Arch. Rat. Mech. Anal.} {\petitbf 87} 
(1985), 107-165 

\ref{52} M. Golubitsky, D. Schaeffer and I. Stewart, {\petitit
Singularities and groups in bifurcation  theory - vol. II}, Springer (New
York) 1988

\ref{53} G. Gaeta, \tit{Bifurcation and symmetry breaking}, {\petitit Phys. Rep.}
{\petitbf 189} (1990), 1-87

\ref{54} P. Chossat and M. Koenig, \tit{Characterization of bifurcations for vector
fields which are equivariant under the action of a compact Lie group}, {\petitit C. R.
Acad. Sci. (Paris)} {\petitbf 318} (1994), 31-36

\ref{55} M. Field, \tit{Equivariant dynamical systems}, {\petitit Bull. A.M.S.}
{\petitbf 76} (1970), 1314-1318

\ref{56} D. Ruelle, \tit{Bifurcations in the presence of a symmetry group}, 
{\petitit Arch. Rat. Mech. Anal.} {\petitbf 51} 
(1973), 136-152

\ref{57} D.H. Sattinger, {\petitit Group theoretic methods in bifurcation theory}, 
{\petitit Lect. Notes Math.} {\petitbf 762},
Springer 1979

\ref{58} D.H. Sattinger, {\petitit Branching in the presence of symmetry}, 
SIAM (Philadelphia) 1984   

\ref{59} P.J. Olver, {\petitit Applications of Lie groups to differential
equations}, Berlin, Springer 1986

\ref{60} G.W. Bluman and S. Kumei, {\petitit Symmetries and
differential equations}, Springer, Berlin 1989

\ref{61} H. Stephani,
{\petitit Differential Equations. Their solution using symmetry}; Cambridge
1991

\ref{62} G. Gaeta, {\petitit Nonlinear symmetry and nonlinear equations},
Kluwer, Dordrecht 1994

\ref{63} P.J. Olver, {\petitit Equivariants, invariants, and symmetry}, Cambridge 1995

\ref{64} G. Gaeta and P. Morando, \tit{Commuting-flow symmetries and common
solutions to differential equations with common symmetry}; Preprint 
{\tt mp-arc 96-618} 1996

\ref{65} R.S. Palais, \tit{The principle of symmetric criticality}, 
\CMP {\petitbf 69} (1979), 19-30

\ref{66} R.S. Palais, \tit{Applications of the symmetric criticality principle in
mathematical physics and differential geometry}, in ``Proceedings of the 1981 Shanghai
symposium on differential geometry and differential equations'', Gu Chaohao ed.,
Science Press, Beijing, 1984

\ref{67} A. Ambrosetti, {\petitit Critical points and nonlinear variational problems},
{\petitit Memoires Soc. Math. France} {\petitbf 49} (supplement to 
{\petitit Bull. S.M.F.} {\petitbf 120}), S.M.F., Paris, 1992

\ref{68} R.S. Palais, \tit{Morse theory on Hilbert manifolds}, {\petitit Topology}
{\petitbf 2} (1963), 299-340

\ref{69} M. Goresky and R. MacPherson, {\petitit Stratified Morse theory}, Springer
1988

\ref{70} G. Gaeta, \tit{Counting symmetry breaking solutions to
symmetric variational problems}, {\petitit Int. J. Theor. Phys.}
{\petitbf 35} (1996), 217-229 

\ref{71} L. Michel, \tit{Extrema des fonctions sur la zone de Brillouin,
invariantes par le groupe de sym\'etrie du cristal et le renversement du
temps}, {\petitit C. R. Acad. Sci. Paris} {\petitbf B-322} (1996), 223-230

\ref{72} N. Cabibbo and L. Maiani, \tit{Weak interactions and the breaking of hadronic
symmetry}; in M. Conversi (ed.), \tit{Evolution of particle Physics (E. Amaldi
Festschrift)}, pp. 50-80, Academic Press 1970

\ref{73} S. Klimek, W. Kondracki, W. Oledzki and P. Sadowski, 
\tit{The density problem for infinite dimensional group
actions}, {\petitit Comp. Math.} {\petitbf 68} (1988), 3-10

\ref{74} O. Babelon and C.M. Viallet, \tit{The riemannian geometry of the configuration
space of gauge theories}, \CMP {\petitbf 81} (1981), 515-525

\ref{75} P. Cotta-Ramusino and C. Reina, \tit{The action of the group of bundle
automorphisms on the space of connections and the geometry of gauge theories},
{\petitit J. Geom. Phys.} {\petitbf 1} (1984), 121-155

\ref{76} M.C. Abbati, R. Cirelli and A. Mani\`a, \tit{The orbit space of the action of
gauge transformation group on connections}, {\petitit J. Geom. Phys.} {\petitbf 6}
(1989), 537-558

\ref{77} M.C. Abbati, R. Cirelli, A. Mani\`a and P. Michor, \tit{The Lie group of
automorphisms of a principal bundle}, {\petitit J. Geom. Phys.} {\petitbf 6} (1989),
215-235

\ref{78} J.J. Duistermaat and G.J. Heckman, \tit{On the variation in the cohomology in
the symplectic form of the reduced phase space}, {\petitit Invent. Math.} {\petitbf
69} (1982), 259

\ref{79} J.M. Bismut, \tit{Localization formulas, superconnections, and the index
theorem for families}, \CMP {\petitbf 103} (1986), 127

\ref{80} L.C. Jeffrey and F.C. Kirwan, \tit{Localization and the quantization
conjecture}, {\petitit Topology} {\petitbf 36} (1997), 647-694

\ref{81} C. Ehresman, \tit{Les prolongements d'une variet\'e differentiable, I-V},
{\petitit C. R. Acad. Sci. (Paris)} {\petitbf 233} (1951), 598,
{\petitbf 233} (1951), 777, {\petitbf 233} (1951), 1081,
{\petitbf 234} (1952), 1028, {\petitbf 234} (1952), 1424

\ref{82} J.L. Koszul, {\petitit Lectures on fibre bundles and differential geometry} 
(Tata Institute), Springer 1986

\bye